\documentclass[reprint,english,prx]{revtex4-1}
\usepackage{hyperref}
\usepackage[hyphenbreaks]{breakurl}
\usepackage[T1]{fontenc}
\usepackage[utf8]{inputenc}
\usepackage{amsmath}
\usepackage{graphicx}
\usepackage{color}
\usepackage[normalem]{ulem}


\usepackage{babel}
\begin{document}

\title{A physical model for dementia}

\author{O. Sotolongo-Costa}
\affiliation{CInC-(IICBA), Universidad Autónoma del Estado de Morelos, 
62209 Cuernavaca, Morelos, México}

\author{L. M. Gaggero-Sager}
\affiliation{CIICAP-(IICBA), Universidad Autónoma del Estado de Morelos, 
62209 Cuernavaca, Morelos, México}

\author{J. T. Becker}
\affiliation{Department of Psychiatry, Department of Neurology and 
Department of Psychology, School of Medicine, University of Pittsburgh, 
Pittsburgh PA 15213, USA}

\author{F. Maestu}
\affiliation{Laboratory of Cognitive and Computational Neuroscience 
(UCM-UPM), Centre for Biomedical Technology (CTB), 
Campus de Montegancedo s/n, Pozuelo de Alarcón, 28223, Madrid, Spain}

\author{O. Sotolongo-Grau}
\affiliation{Alzheimer Research Center and Memory Clinic, Fundació ACE, Institut
Català de Neurociències Aplicades, 08029 Barcelona, Spain}

\collaboration{for the Alzheimer’s Disease Neuroimaging Initiative}
\thanks{Data used in preparation of this article were obtained from the Alzheimer’s Disease Neuroimaging Initiative
(ADNI) database (adni.loni.usc.edu). As such, the investigators within the ADNI contributed to the design
and implementation of ADNI and/or provided data but did not participate in analysis or writing of this report.
A complete listing of ADNI investigators can be found at:
\burl{http://adni.loni.usc.edu/wp-content/uploads/how\_to\_apply/ADNI\_Acknowledgement\_List.pdf}}
\noaffiliation{}


\begin{abstract}
Aging associated brain decline often result in some kind of dementia. 
Even when this is a complex brain disorder a physical model can be used 
in order to describe its general behavior.  This model is based in first 
principles.  A probabilistic model for the development of dementia is 
obtained and fitted to some experimental data obtained from the 
Alzheimer's Disease Neuroimaging Initiative. It is explained how dementia 
appears as a consequence of aging and why it is irreversible.
\end{abstract}
\maketitle

\section{Introduction}

Dementia is a decline in mental ability, caused by damage to brain
cells, that interferes with daily life. Activities of daily living are
usually divided into basic and instrumental activities of daily living
(IADL) \cite{pmid25109674, pmid17521485}. 

Several criteria and methods have been developed as measuring tools 
to implement treatments and diagnoses \cite{pmid15158578, pmid7069156, pmid11156757, pmid21778725}. 
Despite the efforts developed in this field, the relationship between 
IADL performance and mental activity is nowadays implemented using 
only simple statistical approaches like Pearson's or Spearman's correlations.

On the other hand, catastrophe theory, particularly cusp catastrophe 
models, have been used to describe several psychological processes 
and human activities (drinking, sexual interactions, nursing turnover, etc)
\cite{pmid24785249, flay1978catastrophe, BS:BS3830270305, pmid2930621, 
BJOP:BJOP2391, van1992stagewise, pmid11317984, Clair1998, pmid16884652, 
pmid17355897, pmid18765073, pmid20887691, wagner2010predicting, pmid23735493}.
However, in those studies, the data were fit to a cusp surface without 
support from any phenomenological model, so that the physical reasons 
of those processes remain obscure. Here, we introduce a physical 
representation of brain functions representing the brain tasks as 
creation of networks between several neurons.


\section{Theoretical Model}

In order to support a brain task, a network between several neurons is 
created. This network is characterized by a correlation length, $x$,
\cite{Barabasi.2003} that depends both on the topology and on the 
functionality of the network \cite{PhysRevE.80.046104}. The degree of metabolic
activity  necessary to support the task (and the network) is proportional
to the volume of the network determined by
this correlation length. This metabolic activity
is equal to the energy used to maintain the function of the neurons and their links,
$m_{0}$, plus the energy required for the dynamic formation of the specific
network, $m_{x}$.

However each brain task is not instantiated in its own isolated network.
Networks are shared between tasks resulting in connectivity
hubs \cite{pmid24860458}. When several cognitive processes share the same network, 
they may do so without a proportional increase in metabolic demand. 
In order to characterize this phenomenon, we introduce the concept of
synaptic overlap. The degree of synaptic overlap is proportional to 
the mean shared area, which is energized by other processes along the network's correlation
length. This characteristic network overlap has been well described and is
often referred to as a network of networks \cite{Barabasi.2003}. 

So, the energetic balance of the network is summarized as,
\begin{equation}
m_{x}+m_{0}=ax{}^{3}-br^{2}x,\label{eq:model0}
\end{equation}
where $a$ and $b$ are coefficients that convert the geometric characterization
of the network into energy units and $r$ characterizes the synaptic overlap. 
Equation \eqref{eq:model0} describes the possible values of
the system in the space determined by metabolic energy, synaptic overlap, 
and correlation length $(m_{x},r^{2},x)$.

Since neuronal network set up is a synchronized response to an electrical
stimulus \cite{haken2010brain} it seems reasonable that a faster network
configuration involves more energy.
Let us now assume that the metabolic energy for a cognitive task
is proportional to the change rate of the correlation length between neurons,
that is, $m_{x}\sim dx/dt$. So, equation \eqref{eq:model0} could
be written as, 
\begin{equation}
\frac{dx}{dt}=x^{3}-\beta x-\alpha=\frac{\partial U}{\partial x},\label{eq:model1}
\end{equation}
where now $\alpha$ and $\beta$ are functions derived from equation
\eqref{eq:model0} that depend, in general, on metabolic energy, synaptic
overlap and time, and where $U$ is a potential function that corresponds
to the Riemann-Hugoniot surface $x^{3}-\beta x-\alpha=0$ for 
different $\alpha$ and $\beta$ values. Equation
\eqref{eq:model1}, or the equivalent potential, describes a cusp model
\cite{Arnold:1992:CT,Zeeman:1976:CT} that predicts sudden changes
for $x$ values; here $\alpha$ and $\beta$ are known as asymmetry
control parameter and bifurcation control parameter respectively. 

Equation \eqref{eq:model1} is thus  a deterministic model
that relates the energy in a cognitive task network to its correlation length.
However, the brain networks are subject to a high level of noise \cite{haken2010brain}.
The coupling of  millions of neurons in a network in order 
to do a task is necessarily subject to random variations.
In order to apply this model to real data a probabilistic term 
should be added to the model. 

This casts equation \eqref{eq:model1} into a stochastic differential equation,
\begin{equation}
\frac{dx}{dt}=\frac{\partial U}{\partial x}+\sigma\left(x\right)W\left(t\right),\label{eq:sde}
\end{equation}
where $\sigma\left(x\right)$ represents a diffusion process, that will
be assumed to be constant, and which $W\left(t\right)$ is a white
noise Wiener process. Notice that \eqref{eq:sde} is a Langevin equation
where the correlation length corresponds to the position of the particle under
the potential. The corresponding Fokker-Planck equation for the probability
density can be written as,
\begin{equation}
\frac{\partial\rho\left(x,t\right)}{\partial t}=\frac{\partial}{\partial x}\left[\frac{\partial U\left(x\right)}{\partial x}\rho\left(x,t\right)\right]+\sigma\frac{\partial^{2}}{\partial x^{2}}\rho\left(x,t\right).\label{eq:fokker-planck}
\end{equation}
Nevertheless, equation \eqref{eq:fokker-planck} involves two different 
characteristic times. Changes in $x$ occur in the time 
of task processing and brain network assembling, that is, in seconds or minutes. 
Alterations of $U$, and consequently of $\rho$, are due to the 
development of the neurodegenerative diseases that act in a time scale of years.
Since the variation of $x$ in time is faster than the change of $U$,
it can be assumed that $\rho$ changes very slowly over time, and consequently
$\partial\rho/\partial t\simeq0$. 

From this, it is straightforward that,
\begin{equation}
\rho\left(x\right)=Ce^{-\left[\frac{1}{4}x^{4}-\frac{1}{2}\beta x^{2}-\alpha x\right]},\label{eq:prob_solution}
\end{equation}
where $C$ is a normalization constant.

This last expression gives the probability density of obtaining a network
of size $x$ for the steady state case, that is, if the system varies
slowly over time.

Since the probability density for a network with correlation length $x$ 
is known, the entropy of the set of networks,
can be calculated as,
\begin{equation}
	S\left(\alpha, \beta\right) = -\int_{0}^{\infty}{\rho\left(y\right) \mathrm{ln}\left(\rho\left(y\right) \right)dy},\label{eq:entropy}
\end{equation}
showing the natural evolution of the system.

\section{Data fitting}

In order to evaluate our model we fit some real data. 
We should determine first how to model the correlation length 
of the network. On one hand, neurodegenerative diseases affect first the 
largest networks \cite{pmid19376066} and this is reflected in the impairment
of the more complex task.

On the other hand, we should consider the evolution of the brain.
From one organism to other, brain has growth in size and complexity. While
the new evolved life forms are able to learn more complex task, their brain
grow in new layers and connected networks \cite{pmid24210963}.

Notice also that high frequency activity in brain has been associated
to cognitive process \cite{pmid16467513} implying that high functioning 
requires more energy.

So, the correlation
length of the network will be modeled as proportional to the network output.
That is, a bigger network is assumed as needed in order to accomplish  a
more difficult task.

Data used in the preparation of this article were obtained from the Alzheimer's Disease Neuroimaging
Initiative (ADNI) database (adni.loni.usc.edu). The ADNI was launched in 2003 as a public-private
partnership, led by Principal Investigator Michael W. Weiner, MD. The primary goal of ADNI has been to
test whether serial magnetic resonance imaging (MRI), positron emission tomography (PET), other
biological markers, and clinical and neuropsychological assessment can be combined to measure the
progression of mild cognitive impairment (MCI) and early Alzheimer’s disease (AD).

ADNI is a global research effort devote to the research of AD. 
The website group clinical, imaging, genetic and biospecimen biomarkers 
from normal aging to dementia stages. The standardized methods for imaging
and biomarker collection and analysis are intended for facilitating a
cohesive research worldwide. ADNI provides the collected information to 
all registered members.

A sample of 1351 subjects was selected from ADNI cohort.
All available data from these individuals gave 
a total of 3025 study visits. We selected for analysis: positron emission
tomography fluorodeoxyglucose (FDG) standard uptake value ratio,
total brain volume (TBV), intracranial volume (ICV), as well as the functional activities 
questionnaire (FAQ) score. This questionnaire is the information obtained 
from caregivers about IADL performance of patients.
For each subject the brain ratio (BR) was calculated as 
the ratio between TBV and ICV.

For each variable to be fitted into the model, FDG, BR and FAQ,
a linear transformation was applied to
the data in order to normalize it to the interval $[0,1]$.
In the case of FAQ values the transformation was applied
in opposite direction. That is, the FAQ score increases as impairment
of IADL increases but the normalized variable decreases as impairment
of IADL increases.

The network output is proposed as proportional
to IADL; the bifurcation ($\beta$) and asymmetry ($\alpha$) control
parameters are proposed as linear functions of the independent variables \cite{R-Cusp-Package}.
That is, 
\begin{equation}
\begin{array}{c}
x=w_{0}+w_{1}f\\
\alpha=a_{0}+a_{1}u+a_{2}v\\
\beta=b_{0}+b_{1}u+b_{2}v
\end{array},\label{eq:fitting_model-1}
\end{equation}
where $f$, $u$ and $v$ stand for the normalized values of FAQ, FDG and BR and 
$w_{i}$, $a_{i}$ and $b_{i}$ are fitting coefficients. 

All the statistical procedures were made using R Statistical Software
\cite{R}. The R package “cusp” calculate the Cobb's pseudo-$R^{2}$ parameter as a measurement 
of the goodness of fit \cite{R-Cusp-Package}.
Cobb's pseudo-$R^{2}$ and Pearson's $R^{2}$ corresponding to the linear model 
$x=c_{0}+c_{1}f+c_{2}u+c_{3}v$
were calculated and compared each other.
The software fits the data to the \emph{standard} cusp model, where 
the bifurcation is centered at $\alpha = 0$, $\beta = 0$ and $x = 0$. 
However, by requiring to \eqref{eq:fitting_model-1} that $w_{0}=a_{0}=b_{0}=0$, as boundary conditions,
the data were fitted to \eqref{eq:sde}.

\section{Results}

The Cobb's correlation coefficient for the ADNI data
was pseudo-$R^{2}=0.68$ that seems to be a much better fit compared to
the Pearson's correlation coefficient $R^{2}=0.35$ of the equivalent
linear model. Fitting coefficients of the Riemann-Hugoniot surface on 
$(\alpha, \beta, x)$ space were
$w_{1}=4.6$, $a_{1}=6.6$, $a_{2}=5.4$, $b_{1}=2.8$ and $b_{2}=0.1$. 

\begin{figure}
\setlength{\unitlength}{0.95\columnwidth}
\includegraphics[width=\unitlength]{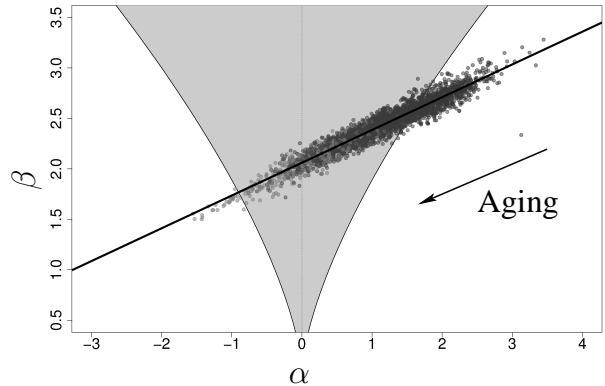}
\caption{\label{fig:plot_bifurcation}Control surface of the cusp model.
Shadowed area represents the bivaluated zone of the cusp. The straight line represents the 
most probable trajectory on the plane. Arrow shows the general direction of aging. Darkness of points
is proportional to the value of the correlation length.}
\end{figure}

\begin{figure}
\setlength{\unitlength}{0.95\columnwidth}
\includegraphics[width=\unitlength]{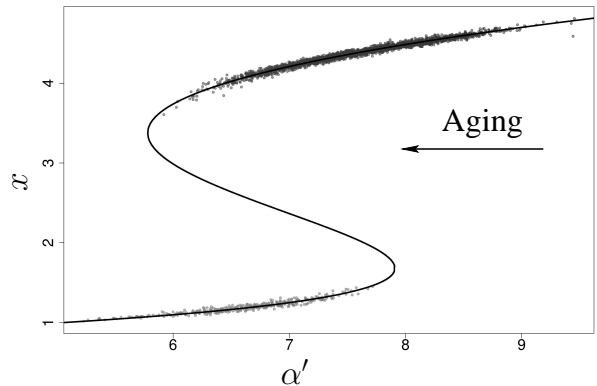}
\caption{\label{fig:plot_transv}{Change of possible values
of the network correlation length along the most probable trajectory 
represented by $\alpha'$ line. The arrow shows the direction of aging.
Darkness of points is proportional to the value of the correlation length.}}
\end{figure}

Figure \ref{fig:plot_bifurcation} shows the control plane of the cusp 
model and how the data distribute for $\alpha$ and $\beta$ values. It can be seen 
that there is a preferential direction along the cusp surface, represented
as a straight line. By translating and rotating the coordinate system, 
so that $\alpha$ lies over this straight line and applying the boundary conditions,
$x$ can be expressed in its "natural" coordinate system ($\alpha'$, $\beta'$). 
Figure \ref{fig:plot_transv} is the representation 
of the data in the new coordinate system.
It shows how the data distribute for $x$ values. 
Here, it can be seen that two different results are possible,
IADL task failure for low values of IADL performance, and success, for high
values of IADL performance.

\begin{figure}
\setlength{\unitlength}{0.95\columnwidth}
\includegraphics[width=\unitlength]{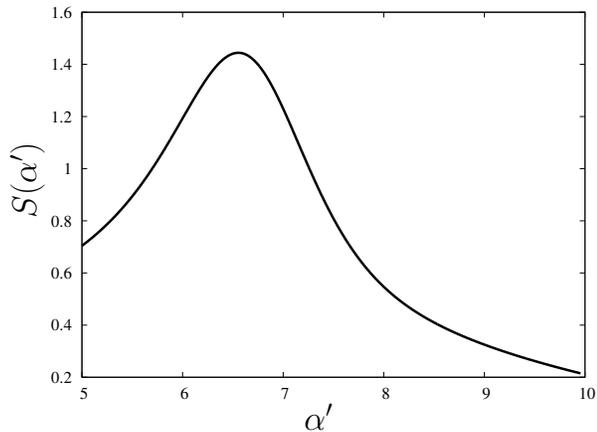}
\caption{\label{fig:entropy}{Entropy of the system as a function
of $\alpha'$. The maximum value of entropy is for 
$\alpha' \simeq{} 6.5$.}}
\end{figure}

\begin{figure}
\setlength{\unitlength}{0.95\columnwidth}
\includegraphics[width=\unitlength]{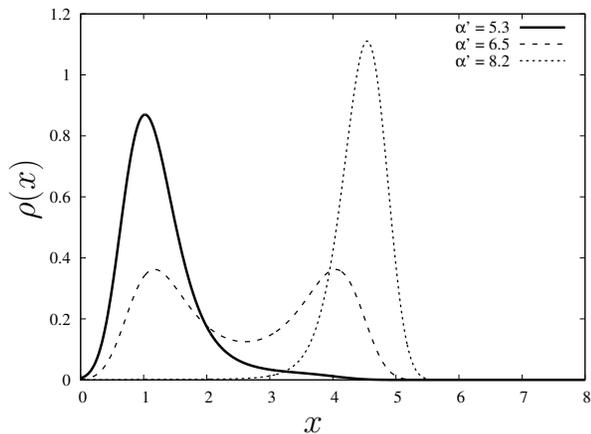}
\caption{\label{fig:probs}{Probability density of obtaining
a network of size $x$ for different values of $\alpha'$ 
along the most probable evolution of the system.}}
\end{figure}

\section{Discussion}

The entropy of the system, calculated according to \eqref{eq:entropy} and represented
in Figure \ref{fig:entropy}, defines the possible evolution of the system in time. 
The system evolves in the general direction of the known aging
processes of brain, represented with arrows in Figures \ref{fig:plot_bifurcation} 
and \ref{fig:plot_transv}. However, the maximum value of entropy 
corresponds to the point of $\alpha = 0$ and $\beta \simeq 2.15$, very close 
to the point where the data intersects the 
$\alpha = 0$  plane ($\beta = 2.06$). 

Older age is generally characterized by decreasing brain volume \cite{pmid9341935}
and a decline in brain glucose metabolism \cite{pmid12482085,pmid17630048,pmid12454908}.
Our model shows that even if this declining process
occurs slowly it could end in a catastrophic failures of IADL, that is, in dementia.
Even when older individuals are more likely to present multiple 
pathologies \cite{pmid26180144} it has been observed that some very old 
people get dementia without the presence of any pathology.
That is, even when age is associated to pathological processes a non small percentage of the 
oldest people get dementia without any pathology \cite{pmid22471863}.

As can be seen in Figure \ref{fig:probs}, the "high energy" states produce only networks 
with high output. However, when there is a loss of brain volume with older age
and a lower energy use, a point is reached where the probability of producing 
a network with a  very low output is not zero. That is, the probability of task 
failure suddenly becomes greater than zero. This probability of failure increases along the 
aging process while the probability of success decreases. At some point along this 
continuum, an individual will be diagnosed with dementia. At very low values of BR
and FDG the probability of success will be zero.

Our results show that functional brain decline is clearly 
observed through the measures of the energy consumption (FDG) and the
brain volume (BR). Dementia progression has been already associated with 
the lesser presence of brain energy consumption \cite{pmid21276856}. 
Furthermore, it has been observed that the decline in energy consumption
increases in advanced disease stages \cite{pmid19660834} pointing to a non
linear relation between both magnitudes. Other authors have linked the IADL 
impairment to brain atrophy \cite{pmid25109674, pmid21646578} and also 
abrupt changes of IADL for different levels of brain atrophy have been 
observed \cite{pmid17521485}, very similar to those changes that our 
model predicts.

However, there is no deterministic
relation between those biomarkers and the onset of dementia. On the
contrary, an individual's decline could follow a random path
through the surface determined by Equation \eqref{eq:model1}. 
Furthermore, the precise moment when the subject falls into dementia
can not be predicted because it is governed by a probability function. 

Beyond statistical inference or linear relationships, 
a few mathematical models link the brain functioning with observed 
measures. However, these models are mainly focused into capturing the patterns 
of the disease instead of offering a general dynamics of the subject 
impairment progression \cite{pmid23450438, pmid24146290, pmid25600871}.
 
Here we offer a general framework that can be used to test the weight 
of clinical variables over the disease. Role of pathological variables
could be easily determined by rewriting $\alpha$ and $\beta$
expression in equations \eqref{eq:fitting_model-1}. The influence of 
comorbidities or other factors usually used as covariables as age or 
genetic factors could be tested the same way. Fitting coefficients should 
show if these variables need to be taken into account. For instance,
it is clear that in equations \eqref{eq:fitting_model-1} the brain atrophy
can be neglected from $\beta$ since the coeffcient $b_{2}$ is an order of 
magnitude smaller than the others. However, the inclusion of
new variables should modulate the trajectory over the surface described 
by \eqref{eq:model1}. So, the research over several variables should 
require much more data in order to show reliable results.

\section{Conclusions}

It has been argued that dementia is the result of a pathological process 
acting on the brain and is fundamentally different from what is called 
healthy aging. However, based on the results of our model, normal aging 
results in a small, but continuous change in the brain that can drive 
loss of performance in IADLs. This presents the provocative possibility 
that at least in some case (e.g., the oldest-old) a dementia syndrome 
could be an end-point of otherwise normal aging. 

However, the solution posed here is only for the steady state.
That is, the curve of Figure \ref{fig:plot_transv} represents the
evolution of system only if changes occur slowly. Stroke, infections, 
and the like cause abrupt changes to the system, and these are not 
accounted for in our model. We do not exclude the possibility that 
dementia could also appear as a consequence of sudden changes 
on the brain. 

While our model explains the general behavior of the data, 
the entropy of the system, shown in Figure \ref{fig:entropy} does not
explain more advanced cases of dementia. This could mean that 
the model should not be applied to the more sparse networks that
would be apparent in demented individuals.

This is a novel approach not only in the field of dementia but more
generally for neurodegenerative diseases. By applying only first 
principles of physics, in this case the laws of thermodynamics, we can 
show how cumulative slow changes in the brain can trigger a catastrophic 
change in the performance of the functional networks. 

\begin{acknowledgments}

This work was supported in part by funds from Fundacio ACE, Institut
Catala de Neurociencies Aplicades, the estate of Trinitat Port-Carbó,
the National Institute on Aging (AG05133) and PRODEP project 
DSA/103.5/15/6986 from SEP, Mexico. 

Data collection and sharing for this project was funded by the Alzheimer's Disease Neuroimaging Initiative
(ADNI) (National Institutes of Health Grant U01 AG024904) and DOD ADNI (Department of Defense award
number W81XWH-12-2-0012). ADNI is funded by the National Institute on Aging, the National Institute of
Biomedical Imaging and Bioengineering, and through generous contributions from the following: AbbVie,
Alzheimer’s Association; Alzheimer’s Drug Discovery Foundation; Araclon Biotech; BioClinica, Inc.; Biogen;
Bristol-Myers Squibb Company; CereSpir, Inc.; Cogstate; Eisai Inc.; Elan Pharmaceuticals, Inc.; Eli Lilly and
Company; EuroImmun; F. Hoffmann-La Roche Ltd and its affiliated company Genentech, Inc.; Fujirebio; GE
Healthcare; IXICO Ltd.; Janssen Alzheimer Immunotherapy Research \& Development, LLC.; Johnson \&
Johnson Pharmaceutical Research \& Development LLC.; Lumosity; Lundbeck; Merck \& Co., Inc.; Meso
Scale Diagnostics, LLC.; NeuroRx Research; Neurotrack Technologies; Novartis Pharmaceuticals
Corporation; Pfizer Inc.; Piramal Imaging; Servier; Takeda Pharmaceutical Company; and Transition
Therapeutics. The Canadian Institutes of Health Research is providing funds to support ADNI clinical sites
in Canada. Private sector contributions are facilitated by the Foundation for the National Institutes of Health
(www.fnih.org). The grantee organization is the Northern California Institute for Research and Education,
and the study is coordinated by the Alzheimer’s Therapeutic Research Institute at the University of Southern
California. ADNI data are disseminated by the Laboratory for Neuro Imaging at the University of Southern
California.

\end{acknowledgments}


\end{document}